\documentclass[aps,showpacs,showkeys,amsmath,amssymb,pra,reprint,superscriptaddress,lengthcheck]{revtex4-1}

\usepackage{dcolumn}
\usepackage{amsmath}
\usepackage{amssymb}
\usepackage{color}
\usepackage{graphicx}
\usepackage{bm}
\usepackage{slashed}

\begin{document}

\title{Generalized excitation of atomic multipole transitions by twisted light modes}

\author{S.~A.-L.~Schulz}
\email{sabrina.schulz@ptb.de}
\affiliation{Physikalisch-Technische Bundesanstalt, D-38116 Braunschweig, Germany}
\affiliation{Technische Universit\"at Braunschweig, D-38106 Braunschweig, Germany}

\author{A.~A.~Peshkov}
\email{a.peshkov@tu-braunschweig.de}
\affiliation{Physikalisch-Technische Bundesanstalt, D-38116 Braunschweig, Germany}
\affiliation{Technische Universit\"at Braunschweig, D-38106 Braunschweig, Germany}

\author{R.~A.~M\"uller}
\affiliation{Physikalisch-Technische Bundesanstalt, D-38116 Braunschweig, Germany}
\affiliation{Technische Universit\"at Braunschweig, D-38106 Braunschweig, Germany}

\author{R.~Lange}
\affiliation{Physikalisch-Technische Bundesanstalt, D-38116 Braunschweig, Germany}

\author{N.~Huntemann}
\affiliation{Physikalisch-Technische Bundesanstalt, D-38116 Braunschweig, Germany}

\author{Chr.~Tamm}
\affiliation{Physikalisch-Technische Bundesanstalt, D-38116 Braunschweig, Germany}

\author{E.~Peik}
\affiliation{Physikalisch-Technische Bundesanstalt, D-38116 Braunschweig, Germany}

\author{A.~Surzhykov}
\affiliation{Physikalisch-Technische Bundesanstalt, D-38116 Braunschweig, Germany}
\affiliation{Technische Universit\"at Braunschweig, D-38106 Braunschweig, Germany} 

\date{\today \\[0.3cm]}

\begin{abstract}

A theoretical study is performed for the excitation of a single atom localized in the center of twisted light modes. Here we present the explicit dependence of excitation rates on critical parameters, such as the polarization of light, its orbital angular momentum projection, and the orientation of its propagation axis with respect to the atomic quantization axis. The effect of a spatial spread of the atom is also considered in detail. The expressions for transition rates obtained in this work can be used for any atom of arbitrary electronic configuration. For definiteness we apply them to the specific case of $^{2}S_{1/2} (F=0) \rightarrow\; ^{2}F_{7/2} (F=3, M=0)$ electric octupole (E3) transition in $^{171}$Yb$^{+}$ ion. Our analytical and numerical results are suitable for the analysis and planning of future experiments on the excitation of electric-dipole-forbidden transitions by twisted light modes in optical atomic clocks.
	
\end{abstract}

\maketitle


\section{\label{sec:Introduction} Introduction}

Since the groundbreaking work of Allen \textit{et al}.\ in 1992 \cite{Allen/PRA:1992}, there has been growing interest in light beams with helical phase fronts. Such twisted (or vortex) beams have many attractive features which can be adapted to the needs of applications \cite{Yao/AOP:2011}. In particular, twisted photons carry a nonzero projection of the orbital angular momentum (OAM) onto the propagation direction and their intensity pattern has an annular character with an intensity minimum in the center \cite{Padgett/PT:2004}. These characteristic features make twisted light a valuable tool for various studies, ranging from the production of coherent superpositions of vortex states in Bose-Einstein condensates \cite{Kapale/PRL:2005} to the direct generation of twisted photons by ultrarelativistic electrons \cite{Chen/PRL:2018, Bogdanov/PRD:2019} or intense vortex harmonics in a plasma \cite{Zhang/NJP:2016}.

In recent years, much interest has been placed on the interaction of twisted light beams with single trapped atoms \cite{Surzhykov/PRA:2015, Quinteiro/PRL:2017, Schulz/PRA:2019}. A remarkable experiment by Schmiegelow and co-workers \cite{Schmiegelow/NC:2016} showed that if atoms are positioned near the low-intensity center of the beam, the twisted light can modify the selection rules and efficiently induce higher-order multipole transitions, in agreement with theoretical predictions \cite{Peshkov/PRA:2017, Afanasev/NJP:2018}. These effects have been demonstrated on the electric quadrupole excitation of $^{40}$Ca$^{+}$ ions, resulting in characteristic population of magnetic sublevels. Moreover, the use of twisted light in this experiment has been shown to result in strong suppression of the AC-Stark shift in the dark beam center. This suppression together with the efficient excitation of non-dipole channels makes twisted light promising for studying and employing atomic clock transitions.

The application of twisted beams to the excitation of single trapped atoms requires a detailed theoretical analysis of the experimental geometry. A typical setup includes a trapped atom whose quantization axis is defined by an applied magnetic field. The atom interacts with a beam of light that is characterized by its polarization and propagation direction. Excitation rates are then strongly affected by the orientation of the atomic quantization axis relative to the propagation direction. Moreover, the strength of atomic transitions may be sensitive not only to the polarization state of the incident light, but also to the spatial spread of atomic position near the beam axis. As a first step towards analyzing such geometrical effects for perfectly localized atoms, recent work \cite{Solyanik-Gorgone/JOSAB:2019} considered a number of forbidden transitions induced by linearly polarized light.

In the present study, we lay down a general theoretical framework for any atom of arbitrary electronic configuration and for various polarization states of light. In addition, a realistic experimental scenario of atoms localized with a finite spatial spread is taken into account. Before discussing our approach for twisted light, in Sec.~\ref{sec:plane} we briefly recall the basic equations governing the excitation of atoms by conventional plane-wave radiation. Here we derive the rates for transitions between magnetic hyperfine sublevels and show that they are sensitive to the polarization of the light beam and to its orientation with respect to the atomic quantization axis. Later in Sec.~\ref{sec:twisted} we discuss the optical excitation by twisted light for the specific case of paraxial Bessel beams. General expressions are derived for the excitation of hyperfine transitions of a particular multipolarity by linearly, radially and azimuthally polarized beams. While the resulting formulas can be applied to an arbitrary atom or ion, in Sec.~\ref{sec:results} we consider the $^{2}S_{1/2} (F=0) \rightarrow\; ^{2}F_{7/2} (F=3)$ electric octupole (E3) transition in $^{171}$Yb$^{+}$, receiving much interest as a candidate for atomic clocks \cite{Godun/PRL:2014, Sanner/N:2019}. Based on our calculations, we show that by choosing proper polarization states of incident twisted light, the transition rate can be significantly enhanced under certain orientations of the applied magnetic field when the atom is placed in the center of the beam. Similar enhancement can also be seen for the ion spatial spread of a few tens of nm. Sec.~\ref{sec:summary} provides concluding remarks.

Hartree atomic units ($\hbar = e = m_{e} =1,\; c=1/ \alpha$) are used throughout the paper.

\section{\label{sec:theory}Theory}
\subsection{\label{sec:plane}Excitation by plane-wave photons}
\subsubsection{\label{sec:plane amplitude}Transition amplitude}

We begin by considering a single trapped atom exposed to a static magnetic field $\bm{B}$ determining the quantization $z_{\text{atom}}$ axis of the atom. Moreover, it is assumed that the atomic nucleus has nonzero spin $\bm{I}$, and hence atomic states $| \alpha F M \rangle$ are characterized by the total angular momentum $\bm{F} = \bm{I} + \bm{J}$, its projection $M$ on the atomic quantization axis, the total electron angular momentum $J$, and all additional quantum numbers $\alpha$. We focus on the atomic excitation process $| \alpha_{i} F_{i} M_{i} \rangle + \gamma \rightarrow | \alpha_{f} F_{f} M_{f} \rangle$ driven by a light field propagating at the angle $\theta$ with respect to the magnetic field. It is well known that all the properties of the excitation process can be traced back to the evaluation of the transition amplitude of the form \cite{Johnson:2007}
\begin{align}
\label{eq:amp_pl_1}
    \mathcal{M}_{M_f M_i}^{(\text{pl})} = \Bigg \langle \alpha_{f} F_{f} M_{f} \Bigg| \sum_{q} \bm{\alpha}_{q} \, \bm{A}_{\lambda}^{(\text{pl})} (\bm{r}_{q}) \Bigg| \alpha_{i} F_{i} M_{i} \Bigg \rangle \, .
\end{align}
Here $q$ runs over all electrons in an atom, and $\bm{\alpha}_q$ denotes the vector of Dirac matrices for the $q$th particle. The incident field is assumed to be a circularly polarized plane wave with helicity $\lambda = \pm 1$. In the Coulomb gauge, the vector potential of this plane wave is given by 
\begin{align}
\label{eq:vec_pl_1}
    \bm{A}_{\lambda}^{(\text{pl})}  (\bm{r}) = \bm{e}_{\bm{k} \lambda} \, e^{i \bm{k} \bm{r}} \, ,
\end{align}
with the polarization vector $\bm{e}_{\bm{k} \lambda}$ and frequency $\omega = kc$. 

In order to evaluate the transition amplitude \eqref{eq:amp_pl_1}, it is convenient to use the multipole expansion of this vector potential  
\begin{align}
\label{eq:vec_pl_2}
        \bm{e}_{\bm{k} \lambda} \, e^{i \bm{k} \bm{r}}  =& \sqrt{2 \pi} \sum_{L M} \sum_{p = 0, 1} i^{L}[L]^{1/2} \, (i \lambda)^{p} \, \notag \\
    &\times D^{L}_{M \lambda} (\phi_k, \theta_k, 0) \bm{a}^{p}_{L M} (\bm{r}) \, ,
\end{align}
where $[L] = 2L + 1$ and $\bm{a}^{p}_{L M} (\bm{r})$ stands for magnetic $(p=0)$ or electric $(p=1)$ multipole components \cite{Rose:1957}. In addition, $D^{L}_{M \lambda}$ is the Wigner $D$-function which depends on the light propagation direction $\hat{\bm{k}} = \bm{k} / k = (\theta_k, \phi_k)$ with respect to the quantization $z$ axis of the entire system, $\theta_k$ and $\phi_k$ being the polar and azimuthal angles. For the further analysis, it is very important to choose this quantization axis in a convenient manner. When the light propagates not along the magnetic field, one can take either $\bm{B}$ or $\bm{k}$ vectors as the quantization $z$ axis of the entire system ``atom $+$ light''. Although it is clear that the observables are independent of a particular choice of coordinate system, we choose here the light propagation ($z_{\text{light}}$) direction as the quantization $z$ axis, so that $\theta_k = \phi_k = 0^{\circ}$. This allows us to simplify Eq.~\eqref{eq:vec_pl_2} by using the relation $D^{L}_{M \lambda} (0, 0, 0) = \delta_{M \lambda}$. On the other hand, we need to perform a rotation of the atomic states $| \alpha_{i} F_{i} M_{i} \rangle$ and $| \alpha_{f} F_{f} M_{f} \rangle$ originally defined in the coordinate system with the quantization $z_{\text{atom}}$ axis oriented along the magnetic field. Such a general transformation
\begin{align}
\label{eq:rotation}
    | \alpha F M \rangle_{\text{atom}} = \sum_{M'} d^{F}_{M' M} (\theta) |  \alpha F M' \rangle_{\text{light}} \, 
\end{align}
can be performed by using the Wigner (small) $d$ functions \cite{Balashov:2000}. In this expression the state vector $| \alpha F M \rangle_{\text{atom}}$ describes the state quantized along the $z_{\text{atom}}$ axis, while $| \alpha F M' \rangle_{\text{light}}$ describes the state quantized along the $z_{\text{light}}$ axis. By using Eq.~\eqref{eq:rotation}, we can rewrite the transition matrix element \eqref{eq:amp_pl_1} as
\begin{align}
\label{eq:amp_pl_1p}
    \mathcal{M}_{M_f M_i}^{(\text{pl})} =& \sum_{M'_i M'_f} d^{F_i}_{M'_i M_i} (\theta) d^{F_f}_{M'_f M_f} (\theta) \, \notag \\
    &\times \Bigg \langle \alpha_{f} F_{f} M'_{f} \Bigg| \sum_{q} \bm{\alpha}_{q} \, \bm{A}_{\lambda}^{(\text{pl})} (\bm{r}_{q}) \Bigg| \alpha_{i} F_{i} M'_{i} \Bigg \rangle \, ,
\end{align}
where all atomic and photonic states on the right-hand side are already defined with respect to the light quantization $z_{\text{light}}$ axis. 

To proceed further, we assume that the electromagnetic field interacts only with atomic electrons and does not affect the nuclear spin. Therefore, it is natural to write the hyperfine wave functions $| \alpha F M' \rangle$ as a linear combination of the corresponding atomic $| \alpha J M'_J \rangle$ and nuclear $|  I M'_I \rangle$ states \cite{Johnson:2007}
\begin{align}
\label{eq:nuclear_atomic}
    | \alpha F M' \rangle = \sum_{M'_J M'_I} \langle J M'_J, \, I M'_I | F M' \rangle | \alpha J M'_J \rangle |  I M'_I \rangle \, .
\end{align}
If we substitute this expression for the atomic states into Eq.~\eqref{eq:amp_pl_1p} and make use of the multipole expansion \eqref{eq:vec_pl_2} together with the Wigner-Eckart theorem, we find after some angular momentum algebra that the transition amplitude reduces to
\begin{align}
\label{eq:amp_pl_2}
    \mathcal{M}_{M_f M_i}^{(\text{pl})} =& \sum_{L p} C_{\alpha_i J_i \alpha_f J_f}^{F_i F_f I} (p L) \,\, (i \lambda)^{p} \, \notag \\
    &\times d^{L}_{\lambda \, \Delta M} (\theta) \, \langle F_i M_i, \, L \Delta M | F_f M_f \rangle \, ,
\end{align}
where $\Delta M = M_f - M_i$ is the difference of the two angular momentum projections on $z_{\text{atom}}$ axis. Here the Clebsch-Gordan coefficients along with Wigner $d$ functions describe the geometry of the experiment, whereas the factors $C_{\alpha_i J_i \alpha_f J_f}^{F_i F_f I} (p L)$ contain information about the coupling to a particular multipole of the electromagnetic field and are defined by
\begin{align}
\label{eq:factor}
    C_{\alpha_i J_i \alpha_f J_f}^{F_i F_f I} &(p L) = \sqrt{2 \pi} \, i^{L} \, [L, F_{i}]^{1/2} \, (-1)^{J_{f} + I + F_{i} + L} \, \notag \\ 
    &\times \left\{ \begin{array}{ccc}
    F_f & F_i & L \\
    J_i & J_f & I
    \end{array} \right\}  
    \langle \alpha_f J_f || H_{\gamma} (pL) || \alpha_i J_i \rangle \, .
\end{align}
Here, we have also introduced the following notation for the the reduced matrix element
\begin{align}
\label{eq:reduced}
    \langle \alpha_f J_f || H_{\gamma} (pL) || \alpha_i J_i \rangle = \Bigg \langle \alpha_{f} J_{f} \Bigg| \Bigg| \sum_{q} \bm{\alpha}_{q} \, \bm{a}_{L, q}^{p} \Bigg| \Bigg| \alpha_{i} J_{i} \Bigg \rangle \, ,
\end{align}
which depends on the specific wave functions of the states.

As can be seen from Eq.~\eqref{eq:amp_pl_2}, the transition amplitude includes a sum over all multipole $(p L)$ contributions. In atomic spectroscopy, however, we can usually restrict ourselves to only a leading multipole term allowed by the angular momentum and parity selection rules. Thus the transition matrix element \eqref{eq:amp_pl_2} is greatly simplified and can be written as
\begin{align}
\label{eq:amp_pl_3}
    \mathcal{M}_{M_f M_i}^{(\text{pl})} =& \, C_{\alpha_i J_i \alpha_f J_f}^{F_i F_f I} (p L) \,\, (i \lambda)^{p} \, \notag \\
    &\times d^{L}_{\lambda \, \Delta M} (\theta) \, \langle F_i M_i, \, L \Delta M | F_f M_f \rangle \, .
\end{align}
This is a very good approximation for light neutral atoms or ions with small ionic charge of interest.

\subsubsection{\label{sec:plane linear}Transition rate for linearly polarized light}

The formulas \eqref{eq:factor}-\eqref{eq:amp_pl_3} were obtained for photons with definite helicity and therefore can be used to describe the excitation of atoms by circularly polarized light. However, it is also instructive to consider linearly polarized incident light.  Along this way, we first need to construct the vector potential for linearly polarized radiation. For example, we can write a plane wave, which is linearly polarized parallel to the reaction $x$-$z$ plane defined by the vectors $\bm{k}$ and $\bm{B}$, as a superposition of the two circularly polarized waves \eqref{eq:vec_pl_1} according to
\begin{align}
\label{eq:pl_h}
    \bm{A}_{\parallel}^{(\text{pl})}  (\bm{r}) = \frac{1}{\sqrt{2}} \left( \bm{A}_{\lambda = +1}^{(\text{pl})} + \bm{A}_{\lambda = -1}^{(\text{pl})} \right) = \bm{e}_{x}  \, e^{i k z} \, .
\end{align}
Similarly, a plane wave linearly polarized perpendicular to the reaction plane is defined by
\begin{align}
\label{eq:pl_v}
    \bm{A}_{\perp}^{(\text{pl})}  (\bm{r}) = \frac{i}{\sqrt{2}} \left( \bm{A}_{\lambda = -1}^{(\text{pl})} - \bm{A}_{\lambda = +1}^{(\text{pl})} \right) = \bm{e}_{y}  \, e^{i k z} \, .
\end{align}
By using these formulas and the absolute-value-squared matrix elements \eqref{eq:amp_pl_3}, we can then introduce the normalized transition rates for absorption of parallel and perpendicular polarized plane-wave light
\begin{align}
\label{eq:rate_pl_h}
    &W_{\parallel}^{(\text{pl})} (\theta) \, \notag \\
    &= \frac{N}{W^{(\text{pl})}_{\text{tot}}} \! \Bigg| \! \frac{1}{\sqrt{2}} \! \left[ \mathcal{M}_{M_f M_i}^{(\text{pl})} (\lambda \! = \! +1) + \mathcal{M}_{M_f M_i}^{(\text{pl})} (\lambda \! = \! -1) \right] \! \Bigg|^2  \, \notag \\
    &= \frac{[L]}{8} \Big| i^{p} \, d^{L}_{+1 \, \Delta M} (\theta) + (-i)^{p} \, d^{L}_{-1 \, \Delta M} (\theta) \Big|^2 , \,
\end{align}
and
\begin{align}
\label{eq:rate_pl_v}
    &W_{\perp}^{(\text{pl})} (\theta) \, \notag \\
    &= \frac{N}{W^{(\text{pl})}_{\text{tot}}} \! \Bigg| \! \frac{i}{\sqrt{2}} \! \left[ \mathcal{M}_{M_f M_i}^{(\text{pl})} (\lambda \! = \! -1) - \mathcal{M}_{M_f M_i}^{(\text{pl})} (\lambda \! = \! +1) \right] \! \Bigg|^2  \, \notag \\
    &= \frac{[L]}{8} \Big| - i^{p} \, d^{L}_{+1 \, \Delta M} (\theta) + (-i)^{p} \, d^{L}_{-1 \, \Delta M} (\theta) \Big|^2 \, ,
\end{align}
where $N$ is some constant factor and $W^{(\text{pl})}_{\text{tot}}$ denotes the total plane-wave transition rate summed over photon polarizations and integrated over magnetic field angles $\theta$. Therefore, the expressions \eqref{eq:rate_pl_h} and \eqref{eq:rate_pl_v} are defined to be dimensionless and independent of line strengths. The combination of Wigner $d$ functions in the last lines of the expressions for $W_{\parallel}^{(\text{pl})} (\theta)$ and $W_{\perp}^{(\text{pl})} (\theta)$ in turn reflects the geometry of the excitation process. In fact, these formulas can be rewritten in terms of polynomials in the sine and cosine functions of $\theta$. For example, the explicit form of the normalized transition rates reads
\begin{align}
\label{eq:rate_pl_ex}
    W_{\parallel}^{(\text{pl})} (\theta) &= \frac{21}{32} \sin^2 \theta \, (1 - 5 \cos^2 \theta )^2 , \, \notag \\
    W_{\perp}^{(\text{pl})} (\theta) &= 0 , \, 
\end{align}
where we assumed $\Delta M = 0$ for an E3 transition.

\subsection{\label{sec:twisted}Excitation by twisted photons}
\subsubsection{\label{sec:twisted amplitude}Transition amplitude}
\begin{figure}[b!]
\includegraphics[width=0.99\linewidth]{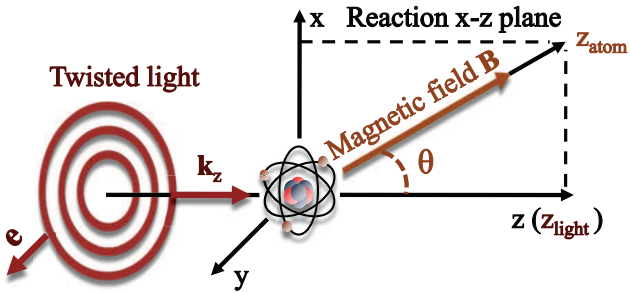}
\caption{Geometry of the optical excitation of a single atom centered on the beam axis ($b=0$) by twisted light with the polarization vector $\bm{e}$. The angle $\theta$ determines the direction of the applied magnetic field $\bm{B}$ defining the quantization $z_{\text{atom}}$ axis of the atom with respect to the light propagation direction taken along the $z_{\text{light}}$ (or $z$) axis.}
\label{fig:geometry}
\end{figure}

Having recalled the theory of atomic excitation by plane waves, we can start to discuss the excitation of atoms by twisted light. In this work, we consider a twisted beam with helicity $\lambda$, the longitudinal component $k_z$ of the linear momentum, the absolute value of the transverse  momentum $ |\bm{k}_{\perp}| = \varkappa$, the photon energy $\omega = c \sqrt{k_z^2 + \varkappa^2}$, and the projection of the total angular momentum (TAM) $m_{\gamma}$. This so-called Bessel state is characterized by the vector potential \cite{Matula/JPB:2013}
\begin{align}
\label{eq:vec_tw_2}
    \bm{A}^{(\text{tw})}  (\bm{r}) = \int a_{\varkappa m_{\gamma}} (\bm{k}_{\perp}) \, \bm{e}_{\bm{k} \lambda} \, e^{i \bm{k} \bm{r}} \, \frac{d^2 \bm{k}_{\perp}}{(2 \pi)^{2}}   \, ,
\end{align}
where the amplitude $a_{\varkappa m_{\gamma}} (\bm{k}_{\perp})$ is given by
\begin{align}
\label{eq:vec_tw_amp}
    a_{\varkappa m_{\gamma}} (\bm{k}_{\perp}) = \frac{2 \pi}{\varkappa} \, (-i)^{m_\gamma} \, e^{i m_\gamma \phi_k} \, \delta (k_{\perp} - \varkappa )  \, .
\end{align}
As seen from Eq.~\eqref{eq:vec_tw_2}, a Bessel beam can be represented as a superposition of plane waves whose wave vectors $\bm{k}$ are uniformly distributed upon the surface of a cone with an opening angle $\theta_{k} = \arctan (\varkappa / k_z)$.

We note that experiments usually deal with the light beams in the so-called paraxial approximation in which the transverse momentum of the photon is much smaller than its longitudinal  momentum, i.e., $\varkappa \ll k_z$, and hence the opening angle $\theta_{k}$ is very small \cite{Matula/JPB:2013}. In this approximation, the vector potential of the Bessel beam \eqref{eq:vec_tw_2} after integration over $\bm{k}_{\perp}$ simplifies to 

\begin{align}
\label{eq:vec_tw_1}
    \bm{A}_{m_l \lambda}^{(\text{tw})}  (\bm{r}) \approx \bm{e}_{\lambda} \, (-i)^{\lambda} J_{m_l}(\varkappa r_{\perp}) \, e^{i m_l \phi} \, e^{i k_z z} \, ,
\end{align}
with $J_{m_l}(\varkappa r_{\perp})$ being the Bessel function of the first kind. Here $m_l = m_\gamma - \lambda$ can be interpreted as the projection of the orbital angular momentum (OAM) of light decoupled from the spin angular momentum (SAM) projection $\lambda$.

As in the case of plane-wave photons \eqref{eq:amp_pl_1}, the interaction between the atoms and the twisted light \eqref{eq:vec_tw_2} can be described by the matrix element
\begin{align}
\label{eq:amp_tw_1}
    \mathcal{M}_{M_f M_i}^{(\text{tw})} &= \Bigg \langle \alpha_{f} F_{f} M_{f} \Bigg| \sum_{q} \bm{\alpha}_{q} \, \bm{A}_{m_l \lambda}^{(\text{tw})} (\bm{r}_{q}) \Bigg| \alpha_{i} F_{i} M_{i} \Bigg \rangle \, \notag \\
    &= \int a_{\varkappa m_{\gamma}} (\bm{k}_{\perp}) \, e^{-i \bm{k}_{\perp} \bm{b}} \, \notag \\
    &\, \times \Bigg \langle \alpha_{f} F_{f} M_{f} \Bigg| \sum_{q} \bm{\alpha}_{q} \, \bm{e}_{\bm{k} \lambda} \, e^{i \bm{k} \bm{r}_{q}} \Bigg| \alpha_{i} F_{i} M_{i} \Bigg \rangle \frac{d^2 \bm{k}_{\perp}}{(2 \pi)^{2}}  \, .
\end{align}
Here the impact parameter $\bm{b} = (b_x, b_y, 0)$ specifies the position of a target atom with regard to the beam axis. The introduction of this parameter is necessary, since the Bessel beam \eqref{eq:vec_tw_1} has an inhomogeneous field distribution in the transverse plane, thereby making the excitation process very sensitive to $\bm{b}$. Calculations of the transition amplitude \eqref{eq:amp_tw_1} can again be simplified through use of the standard multipole expansion of a plane wave \eqref{eq:vec_pl_2}. Then, upon integrating over $k_{\perp}$ with the delta function $\delta (k_{\perp} - \varkappa )$ and making use of the integral representation of the Bessel function \cite{Peshkov/PRA:2017}
\begin{align}
\label{eq:bessel}
    &\int_{0}^{2 \pi} e^{i (m_{\gamma} - M) \phi_k - i \varkappa b \cos (\phi_k - \phi_b)} \, \frac{d \phi_k}{2 \pi} \, \notag \\
    &= (-i)^{m_{\gamma} - M} e^{i (m_{\gamma} - M) \phi_b} J_{m_{\gamma} - M} (\varkappa b) \, ,
\end{align}
we readily find, after some algebraic manipulations similar to those in Eqs.~\eqref{eq:rotation}-\eqref{eq:amp_pl_3}, that
\begin{align}
\label{eq:amp_tw_2}
    &\mathcal{M}_{M_f M_i}^{(\text{tw})} (\bm{b}) = \, C_{\alpha_i J_i \alpha_f J_f}^{F_i F_f I} (p L) \,\, (i \lambda)^{p} \, (-1)^{m_l + \lambda} \, \notag \\
    & \,\,\,\,\,\,\, \times \langle F_i M_i, \, L \Delta M | F_f M_f \rangle \, \sum_{M} i^M \, e^{i (m_l + \lambda - M) \phi_b} \, \notag \\ 
    &\,\,\,\,\,\,\, \times J_{m_l + \lambda - M} (\varkappa b) \, d^{L}_{M \, \lambda} (\theta_k) \, d^{L}_{M \, \Delta M} (\theta) \, ,
\end{align}
where again the $C_{\alpha_i J_i \alpha_f J_f}^{F_i F_f I} (p L)$ is given by Eq.~\eqref{eq:factor} and summation over multipoles $(p L)$ is restricted to the single leading term.

By comparing the transition amplitudes for plane waves \eqref{eq:amp_pl_3} and twisted light \eqref{eq:amp_tw_2}, we observe that the key difference between them arises from the summation over the projection quantum numbers $M$ in $\mathcal{M}_{M_f M_i}^{(\text{tw})}$. This sum contains the Bessel functions $J_{m_l + \lambda - M} (\varkappa b)$ and the exponential factors $\exp [i (m_l + \lambda - M) \phi_b]$ that describe the impact-parameter dependence of the matrix element. It is worth stressing that the transition amplitude \eqref{eq:amp_tw_2} for twisted light can be greatly simplified when the single atom is placed on the beam axis ($b=0$), namely
\begin{align}
\label{eq:amp_tw_3}
    &\mathcal{M}_{M_f M_i}^{(\text{tw})} (b = 0) = \, C_{\alpha_i J_i \alpha_f J_f}^{F_i F_f I} (p L) \,\, (i \lambda)^{p} \, (-i)^{m_l + \lambda} \, \notag \\
    & \,\, \times \langle F_i M_i, \, L \Delta M | F_f M_f \rangle \, d^{L}_{m_l + \lambda , \, \lambda} (\theta_k) \, d^{L}_{m_l + \lambda , \, \Delta M} (\theta) \, .
\end{align}
This equation explicitly expresses the fact that $\mathcal{M}_{M_f M_i}^{(\text{tw})}$ depends on the opening angle $\theta_k$, helicity $\lambda$, and OAM $m_l$ of twisted light, as well as on the angle $\theta$ between the light propagation direction and the external magnetic field [cf.~Fig.~\ref{fig:geometry}].

\subsubsection{\label{sec:twisted linear}Transition rate for linear polarization}

\begin{figure}[t!]
\includegraphics[width=0.99\linewidth]{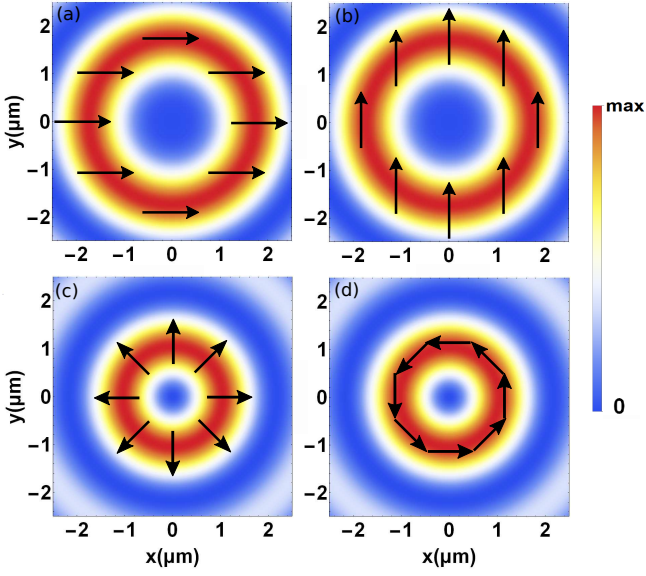}
\caption{Intensity profiles of paraxial Bessel beams of light of wavelength $467$ nm and opening angle $\theta_k = 7.5^{\circ}$. The directions of the electric field are indicated by black arrows: (a) parallel and (b) perpendicular linearly polarized twisted beams with OAM $m_l = +2$; (c) radially polarized beam; (d) azimuthally polarized beam.}
\label{fig:int_pol}
\end{figure}

The formula \eqref{eq:amp_tw_2} shows how to calculate the amplitude for excitation by twisted light with definite helicity $\lambda$. Similar to before, we can again use this amplitude to analyze the absorption of linearly polarized twisted photons. This can be done easily within the paraxial approximation in which the OAM and SAM are independent from each other \cite{Matula/JPB:2013}. In analogy with plane waves, it is possible to construct the vector potential for twisted photons linearly polarized within the reaction $x$-$z$ plane defined by the magnetic field $\bm{B}$ direction and the light propagation $\bm{k}_z$ direction: 

\begin{align}
\label{eq:tw_h}
    \bm{A}_{\parallel}^{(\text{tw})}  (\bm{r}) &= \frac{i}{\sqrt{2}} \left( \bm{A}_{m_l \, \lambda = +1}^{(\text{tw})} - \bm{A}_{m_l \, \lambda = -1}^{(\text{tw})} \right) \, \notag \\ 
    &= \bm{e}_{x}  \, J_{m_l}(\varkappa r_{\perp}) \, e^{i m_l \phi} \, e^{i k_z z} \, ,
\end{align}
while for twisted photons polarized perpendicular to the reaction plane we find
\begin{align}
\label{eq:tw_v}
    \bm{A}_{\perp}^{(\text{tw})}  (\bm{r}) &= \frac{1}{\sqrt{2}} \left( \bm{A}_{m_l \, \lambda = +1}^{(\text{tw})} + \bm{A}_{m_l \, \lambda = -1}^{(\text{tw})} \right) \, \notag \\ 
    &= \bm{e}_{y}  \, J_{m_l}(\varkappa r_{\perp}) \, e^{i m_l \phi} \, e^{i k_z z} \, .
\end{align}
The intensity profile of these beams with an OAM $m_l = +2$ is shown in Figs.~\ref{fig:int_pol} (a) and (b) together with the polarization vectors.

By using the vector potential \eqref{eq:tw_h} and the matrix element \eqref{eq:amp_tw_2}, we find the normalized transition rate for absorption of parallel polarized twisted light:
\begin{align}
\label{eq:rate_tw_h}
    W_{\parallel}^{(\text{tw})} (\theta, \bm{b})  = \frac{N}{W^{(\text{pl})}_{\text{tot}}} \Bigg| & \frac{i}{\sqrt{2}} \left[ \mathcal{M}_{M_f M_i}^{(\text{tw})} (m_l, \lambda \! = \! +1) \, \notag \right. \\ 
    & \left. - \mathcal{M}_{M_f M_i}^{(\text{tw})} (m_l, \lambda \! = \! -1) \right] \Bigg|^2 \, .
\end{align}
On the other hand, for perpendicular polarized twisted light \eqref{eq:tw_v} we have
\begin{align}
\label{eq:rate_tw_v}
    W_{\perp}^{(\text{tw})} (\theta, \bm{b})  = \frac{N}{W^{(\text{pl})}_{\text{tot}}} \Bigg| & \frac{1}{\sqrt{2}} \left[ \mathcal{M}_{M_f M_i}^{(\text{tw})} (m_l, \lambda \! = \! +1) \, \notag \right. \\ 
    & \left. + \mathcal{M}_{M_f M_i}^{(\text{tw})} (m_l, \lambda \! = \! -1) \right] \Bigg|^2 \, .
\end{align}
If the atom is placed on the beam axis ($b=0$), these expressions simplify to
\begin{align}
\label{eq:rate_tw_hv_b0}
    W_{\parallel, \perp}^{(\text{tw})} (\theta, b = 0) &= \frac{[L]}{8} \Big| -i^{p+1} \, d^{L}_{m_l + 1 , \, +1} (\theta_k) \, d^{L}_{m_l + 1 , \, \Delta M} (\theta) \, \notag \\
    & \mp (-i)^{p-1} \, d^{L}_{m_l - 1, \, -1} (\theta_k) \, d^{L}_{m_l - 1, \, \Delta M} (\theta) \Big|^2 \, ,
\end{align}
where we have made use of Eq.~\eqref{eq:amp_tw_3}. Moreover, using the explicit expressions of Wigner $d$ functions, we obtain 
\begin{align}
\label{eq:rate_tw_hv_b0_ex}
    &W_{\parallel, \perp}^{(\text{tw})} (\theta, b = 0) \, \notag \\
    &= \frac{21}{8192} \Big| 5 \sin^2 \theta_k \, (1 + \cos \theta_k) \, \sin^3 \theta \mp (1 - \cos \theta_k) \, \notag \\   
    & \;\;\; \times (1 - 10 \cos \theta_k - 15 \cos^2 \theta_k) \, (1 - 5 \cos^2 \theta) \, \sin \theta  \Big|^2 \, ,
\end{align}
which corresponds to $\Delta M = M_f - M_i = 0$ and $m_l = +2$ for an E3 transition.

\subsubsection{\label{sec:twisted rad_az}Transition rate for radially and azimuthally polarized beams}
With twisted light modes, it is possible to generate richer polarization patterns compared to plane waves \cite{Dorn/PRL:2003, Tidwell/AO:1990}. For example, the so-called radially polarized beam, whose vector potential
\begin{align}
\label{eq:tw_r}
    \bm{A}_{\text{r}}^{(\text{tw})}  (\bm{r}) &= -\frac{i}{\sqrt{2}} \left( \bm{A}_{m_l = -1 \, \lambda = +1}^{(\text{tw})} + \bm{A}_{m_l = +1 \, \lambda = -1}^{(\text{tw})} \right) \, \notag \\ 
    &= \bm{e}_{r}  \, J_{1}(\varkappa r_{\perp}) \, e^{i k_z z} \, 
\end{align}
can be constructed as a linear combination of two twisted beams with helicity $\lambda = \pm 1$ and OAM $m_l = \mp 1$ \cite{Quinteiro/PRA:2017}. The corresponding intensity profile and electric field directions are shown in Fig.~\ref{fig:int_pol} (c). We see that in this case the electric field vector at any point on the beam profile is oriented in the radial direction $\bm{e}_{r}$. With Bessel beams we can also obtain another solution, namely the azimuthally polarized beam defined as
\begin{align}
\label{eq:tw_az}
    \bm{A}_{\text{az}}^{(\text{tw})}  (\bm{r}) &= \frac{1}{\sqrt{2}} \left( \bm{A}_{m_l = +1 \, \lambda = -1}^{(\text{tw})} - \bm{A}_{m_l = -1 \, \lambda = +1}^{(\text{tw})} \right) \, \notag \\ 
    &= \bm{e}_{\phi}  \, J_{1}(\varkappa r_{\perp}) \, e^{i k_z z} \, ,
\end{align}
where the polarization vector $\bm{e}_{\phi}$ implies that its electric field vector is always perpendicular to the radial vector, as depicted in Fig.~\ref{fig:int_pol} (d). We note that both radially and azimuthally polarized beams do not have a well-defined OAM projection $m_l$; they can be seen as a superposition of two twisted modes with different OAM $m_l = \pm 1$ \cite{Quinteiro/PRA:2017}.

Similarly to before, we can use the vector potential \eqref{eq:tw_r} to calculate the transition rate for absorption of radiation of radial polarization for arbitrary impact parameters
\begin{align}
\label{eq:rate_tw_r}
    W_{\text{r}}^{(\text{tw})} (\theta, \bm{b})  = &\frac{N}{W^{(\text{pl})}_{\text{tot}}} \Bigg|  -\frac{i}{\sqrt{2}} \left[ \mathcal{M}_{M_f M_i}^{(\text{tw})} (m_l \! = \! -1, \lambda \! = \! +1) \, \notag \right. \\ 
    & \left. + \mathcal{M}_{M_f M_i}^{(\text{tw})} (m_l \! = \! +1 , \lambda \! = \! -1) \right] \Bigg|^2 \, .
\end{align}
In the case of azimuthal polarization \eqref{eq:tw_az}, we find
\begin{align}
\label{eq:rate_tw_az}
    W_{\text{az}}^{(\text{tw})} (\theta, \bm{b})  = &\frac{N}{W^{(\text{pl})}_{\text{tot}}} \Bigg|  \frac{1}{\sqrt{2}} \left[ \mathcal{M}_{M_f M_i}^{(\text{tw})} (m_l \! = \! +1, \lambda \! = \! -1) \, \notag \right. \\ 
    & \left. - \mathcal{M}_{M_f M_i}^{(\text{tw})} (m_l \! = \! -1 , \lambda \! = \! +1) \right] \Bigg|^2 \, .
\end{align}
Again, these formulas are significantly simplified if one considers the atom with $b = 0$. By substituting Eq.~\eqref{eq:amp_tw_3} into Eqs.~\eqref{eq:rate_tw_r} and \eqref{eq:rate_tw_az}, we obtain the normalized transition rates
\begin{align}
\label{eq:rate_tw_raz_b0}
    &W_{\text{r,az}}^{(\text{tw})} (\theta, b = 0) = \frac{[L]}{8} \Big| d^{L}_{0 \, \Delta M} (\theta) \Big|^2 \,  \notag \\
    &\;\; \times \Big| (-i)^{p} \, d^{L}_{0 \, -1} (\theta_k) \pm i^{p} \, d^{L}_{0 \, +1} (\theta_k) \Big|^2 \, ,
\end{align}
where we have used the fact that both twisted components have the TAM projection $m_{\gamma} = m_l + \lambda = 0$. In particular, the explicit expressions for $W_{\text{r,az}}^{(\text{tw})}$ are
\begin{align}
\label{eq:rate_tw_raz_ex}
    W_{\text{r}}^{(\text{tw})} (\theta, b = 0) =& \frac{21}{128} \sin^2 \theta_k \, (1 - 5 \cos^2 \theta_k)^2 \,  \notag \\
    &\times \cos^2 \theta \, (3 - 5 \cos^2 \theta)^2 \,  , \notag \\ 
    W_{\text{az}}^{(\text{tw})} (\theta, b = 0) =& \, 0 \, ,
\end{align}
assuming that $\Delta M = 0$ and $m_l = +2$ for an E3 transition.

\subsubsection{\label{sec:target}Nonlocalized atoms}
The transition rates for excitation by twisted light have been derived above for an atom whose position with respect to the beam axis is well-defined. Of course, this idealistic scenario is impossible to achieve in a real atom trap experiment. For example, in recent experiments \cite{Afanasev/NJP:2018, Schmiegelow/NC:2016}, a single laser-cooled $^{40}$Ca$^{+}$ ion trapped in a Paul trap had a thermal spatial spread of about $60$ nm. To account for this spatial distribution effect, we introduce the probability to find an atom at the distance $\bm{b}$ from the beam center:
\begin{align}
\label{eq:distr}
    f (\bm{b}) = \frac{1}{2 \pi \sigma^2} e^{-\frac{\bm{b}^2}{2 \sigma^2}} \, .
\end{align}
This Gaussian distribution is assumed to be centered on the beam axis and characterized by the width $\sigma$. The mean transition rates are then obtained by integrating over the impact parameter
\begin{align}
\label{eq:mean}
    W^{(\text{tw})}_{\parallel, \perp, \text{r,az}} (\theta) = \int f (\bm{b}) \, W^{(\text{tw})}_{\parallel, \perp, \text{r,az}} (\theta, \bm{b}) \, d^2 \bm{b} \, .
\end{align}
This formula enables us to understand how the excitation probability depends not only on the mutual orientation of photon and atomic axes, but also on the width $\sigma$ of the atomic spatial distribution.

\subsubsection{\label{sec:lg}Laguerre-Gaussian beams}
\begin{figure}[b!]
\includegraphics[width=0.76\linewidth]{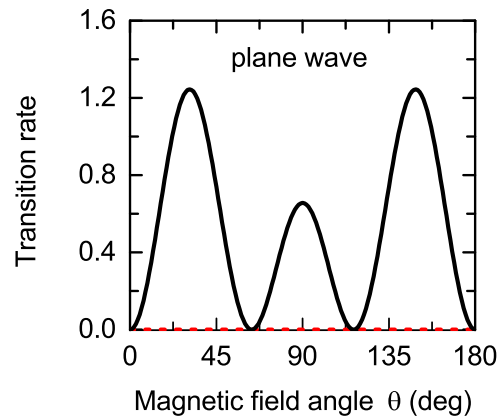}
\caption{Normalized transition rate for absorption of plane-wave light linearly polarized parallel to the reaction plane (black solid line) as a function of tilt angle $\theta$ of the magnetic field for the $^{2}S_{1/2} (F_i = 0, \, M_i = 0) \rightarrow\; ^{2}F_{7/2} (F_f = 3 , \, M_f = 0)$ electric octupole (E3) transition in $^{171}$Yb$^{+}$. The transition rate for perpendicular polarized plane waves (red dashed line) vanishes identically for all angles $\theta$.}
\label{fig:pl}
\end{figure}

In this study, we present the theory of excitation of trapped atoms by Bessel light beams. In some recent experiments, however, another type of twisted light modes, namely Laguerre-Gaussian modes, are employed \cite{Peshkov/PRA:2017, Schmiegelow/NC:2016}. While the detailed discussion of these modes is beyond the scope of our work, we mention here that both Bessel and Laguerre-Gaussian solutions behave like $r_{\perp}^{|m_l|} e^{i m_l \phi}$ near the optical vortex, i.e., for small $r_{\perp}$. This implies that the formulas derived above can be applied to the excitation by Laguerre-Gaussian beams of the same polarization and OAM if the spatial distribution of atoms is smaller than the characteristic size of the first bright ring of the beam, which is of the order of a few microns in our case [cf.~Fig.~\ref{fig:int_pol}].

\section{\label{sec:results}Results and discussion}
In the previous sections, we obtained the formulas which allow us to calculate the (normalized) excitation rates for twisted light and plane waves. While the developed formalism can be applied to any atom or ion, we focus here on the $4f^{14} 6s \; ^{2}S_{1/2} \rightarrow \; 4f^{13} 6s^2 \; ^{2}F_{7/2} $ electric octupole (E3) transition in singly ionized ytterbium. This transition has attracted much experimental and theoretical attention as a candidate for a frequency standard \cite{Sanner/N:2019, Biemont/PRL:1998}. For the specific case of $^{171}$Yb$^{+}$, we consider the initial and final hyperfine levels $F_i = 0$ and $F_f = 3$. A weak magnetic field is applied to split the hyperfine multiplet of the excited state by the Zeeman effect, making the individual $M_f$ sublevels distinguishable. In what follows, we concentrate on the transition between the $M_i = 0$ and $M_f = 0$ magnetic hyperfine sublevels that do not show a linear Zeeman effect.

\begin{figure}[b!]
\includegraphics[width=0.76\linewidth]{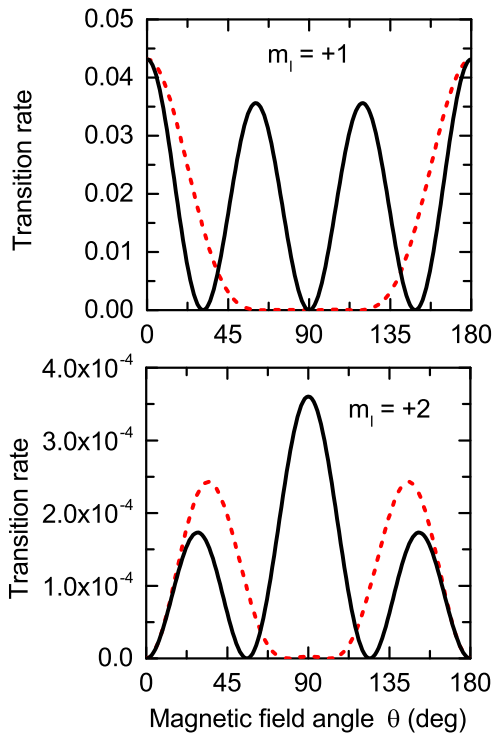}
\caption{Normalized transition rates for absorption of parallel (black solid lines) and perpendicular (red dashed lines) linearly polarized twisted Bessel light with OAM $m_l = +1$ (top panel) and $m_l = +2$ (bottom panel) by a single $^{171}$Yb$^{+}$ ion placed on the beam axis ($b = 0$), while the opening angle is $\theta_k = 7.5^{\circ}$. All other parameters are as in Fig.~\ref{fig:pl}.}
\label{fig:tw_m12_xy}
\end{figure}
\begin{figure}[t!]
\includegraphics[width=0.76\linewidth]{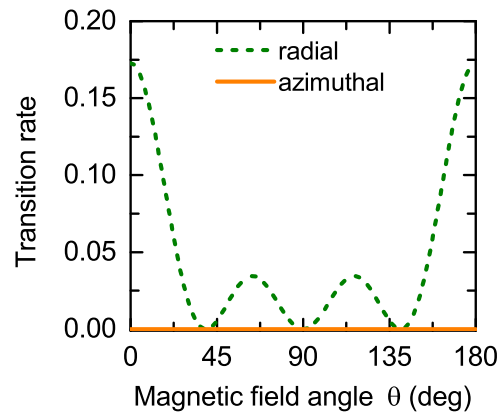}
\caption{The same as Fig.~\ref{fig:tw_m12_xy}, but for the radially polarized Bessel beam (green dashed line). The transition rate for the azimuthally polarized beam (orange solid line) equals zero for all angles $\theta$ if $b = 0$.}
\label{fig:ra}
\end{figure}
\subsection{\label{sec:results_pw}Excitation by plane waves}
Before considering the excitation of an $^{171}$Yb$^{+}$ ion by Bessel beams, we shall briefly review the results for plane waves \cite{Taylor:1996}. Fig.~\ref{fig:pl} shows the transition rates normalized to the factor $W^{\text{(pl)}}_{\text{tot}}$ for the $^{2}S_{1/2} (F_i = 0, \, M_i = 0) \rightarrow\; ^{2}F_{7/2} (F_f = 3 , \, M_f = 0)$ transition induced by plane-wave radiation with a wavelength of $467$ nm. Calculations are performed for the light polarized either parallel (black solid line) or perpendicular (red dashed line) to the reaction plane. As seen from Fig.~\ref{fig:pl}, the transition rates \eqref{eq:rate_pl_h} and \eqref{eq:rate_pl_v} depend strongly on both the tilt angle $\theta$ of the magnetic field and the polarization of light. For example, the excitation rate $W_{\perp}^{(\text{pl})}$ vanishes identically for all angles $\theta$ when the incident light is perpendicular polarized [cf.~Eq.~\eqref{eq:rate_pl_ex}]. In the case of parallel polarization, $W_{\parallel}^{(\text{pl})}$ exhibits oscillations and reaches maximum values at $\theta = 30^{\circ}$, $\theta = 90^{\circ}$, and $\theta = 150^{\circ}$.

\subsection{\label{sec:center}Excitation by twisted light: atoms on the beam axis}

Having examined the excitation by linearly polarized plane waves, we can study the effects of twisted beams on transition probabilities. Again, we focus on paraxial Bessel beams polarized parallel or perpendicular to the reaction plane. Here the atom is assumed to be placed directly on the beam axis, $b = 0$. The corresponding transition rates \eqref{eq:rate_tw_hv_b0} are displayed in Fig.~\ref{fig:tw_m12_xy} for two OAM of incident light $m_l = +1$ (top panel) and $m_l = +2$ (bottom panel). It is apparent that the $\theta$ dependence of $W^{(\text{tw})}$ differs significantly from its plane-wave counterpart. In particular, while $W_{\perp}^{(\text{pl})}$ is identically zero for arbitrary orientation of the magnetic field, this is not the case for $W_{\perp}^{(\text{tw})}$. For example, for OAM projection $m_l = +1$, the normalized transition rate for twisted photons exhibits maximum values of $0.043$ at $\theta = 0^{\circ}$ and $180^{\circ}$. On the other hand, the positions of the absorption peaks for OAM $m_l = +2$ are shifted to $\theta = 35^{\circ}$ and $145^{\circ}$. It is worth stressing that the transition rate for twisted light polarized within the reaction plane $W^{(\text{tw})}_{\parallel}$ is also very sensitive to the OAM projection. For $m_l = +1$ the maxima occur at $\theta = 0^{\circ}$, $60^{\circ}$, $120^{\circ}$, and $180^{\circ}$, while for $m_l = +2$ there are maxima at $\theta = 30^{\circ}$, $90^{\circ}$, and $150^{\circ}$. As seen from Fig.~\ref{fig:tw_m12_xy}, not only the angular pattern, but also the absolute values of transition rates depend on the OAM of light. They are reduced by almost two orders of magnitude when the OAM projection $m_l$ is increased from $+1$ to $+2$. This is caused by the reduction of the intensity of light near the vortex
line as OAM increases \cite{Matula/JPB:2013, Abramowitz:1964}.

So far we have discussed the excitation of atoms by linearly polarized light. In order to further study the polarization dependence of the transition rates, we also consider radially and azimuthally polarized incident beams. The calculations \eqref{eq:rate_tw_raz_b0} for these nontrivial polarization states and for an atom located on the beam axis are displayed in Fig.~\ref{fig:ra}. This figure shows that the maximal excitation probability for radially polarized beam can be achieved by applying the external magnetic field along the light axis. In contrast, no excitation by azimuthally polarized beam \eqref{eq:rate_tw_raz_ex} is observed for $b = 0$. As will be demonstrated in the next section, this suppression is partially removed if the spatial spread of the trapped atom is not negligible with respect to the inverse transverse momentum $\varkappa^{-1}$ of the beam.

\begin{figure}[t!]
\includegraphics[width=0.76\linewidth]{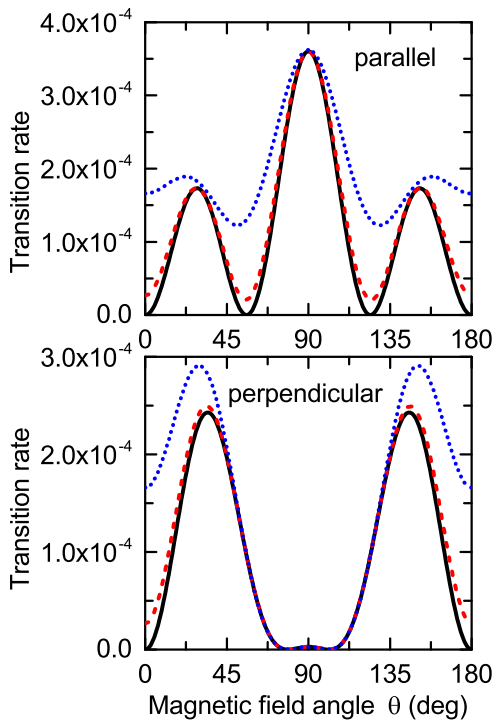}
\caption{Transition rates for absorption of parallel (top panel) and perpendicular (bottom panel) linearly polarized twisted Bessel light with OAM $m_l = +2$ and opening angle $\theta_k = 7.5^{\circ}$ when a single ion is placed on the beam axis (black solid lines), and when the centered ion has a spatial spread of $\sigma = 20$ nm (red dashed lines) or $\sigma = 50$ nm (blue dotted lines) in the beam center. All other parameters are as in Fig.~\ref{fig:pl}.}
\label{fig:av_tw_m2_xy}
\end{figure}
\subsection{\label{sec:distr}Excitation by twisted light: spatial spread of ions}
The above calculations have been carried out for the atom positioned precisely at the beam center, $b=0$. Such an ideal scenario, however, is impossible to attain in real atom-trap experiments in which the atomic spatial spread has been found to be about 60 nm \cite{Schmiegelow/NC:2016}. We make use of Eq.~\eqref{eq:mean} for the Gaussian distribution \eqref{eq:distr} to study the effect of atomic delocalization, and the corresponding transition rates are displayed in Fig.~\ref{fig:av_tw_m2_xy} for Bessel beams linearly polarized parallel and perpendicular the reaction plane. Calculations have been done for OAM $m_l = +2$ and for atomic spatial distributions of width $\sigma = 20$ nm (red dashed lines) or $\sigma = 50$ nm (blue dotted lines) centered on the beam axis. In addition, these results are compared with idealized predictions for single atoms located exactly at the beam center (black solid lines). The atomic delocalization significantly affects the rates $W^{(\text{tw})}_{\parallel}$, as shown in the top panel of Fig.~\ref{fig:av_tw_m2_xy}. For example, the oscillations of the transition rate become less pronounced at $\theta < 60^{\circ}$ and $\theta > 120^{\circ}$. The principal maximum of the transition rate remains unchanged and occurs for the perpendicular orientation of the trap magnetic field. The finite spatial spread of the atom also modifies the transition rates for absorption of Bessel photons polarized perpendicular to the reaction plane. As can be seen from the bottom panel of Fig.~\ref{fig:av_tw_m2_xy}, the rate $W^{(\text{tw})}_{\perp}$ is enhanced at $\theta < 45^{\circ}$ and $\theta > 135^{\circ}$, and the positions of the maxima become slightly shifted.

\begin{figure}[b!]
\includegraphics[width=0.76\linewidth]{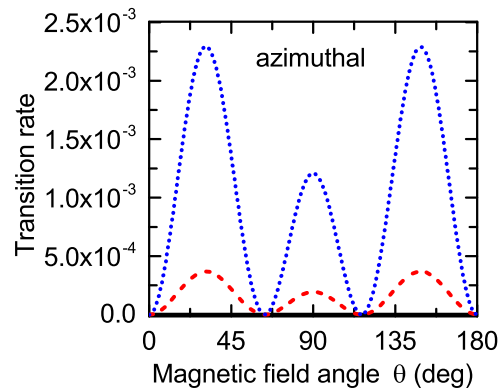}
\caption{The same as Fig.~\ref{fig:av_tw_m2_xy}, but for the azimuthally polarized Bessel beams.}
\label{fig:av_tw_az}
\end{figure}

It should be noted that the spatial delocalization of the atom can strongly affect the transition rates also for other types of polarization. In particular, we see from Fig.~\ref{fig:av_tw_az} that while $W^{(\text{tw})}_{\text{az}}$ for azimuthal polarization vanishes identically when the atom is positioned precisely at $b=0$, the rate of photon absorption increases with the width of the atomic distribution $\sigma$. For example, if $\sigma = 50$ nm, $W^{(\text{tw})}_{\text{az}}$ reaches the maximum value of $2.3 \times 10^{-3}$ at $\theta = 30^{\circ}$ and $150^{\circ}$.

%


%
%
%
\section{\label{sec:summary}Summary and outlook}
In summary, we have performed a theoretical study of the excitation of a single trapped atom by twisted light. Special attention has been paid to the dependence of the excitation rates on the polarization of incident light and on the orientation of the beam axis with respect to the atomic quantization axis, which is defined by the applied magnetic field. In order to investigate this geometrical dependence, we have employed first-order perturbation theory to describe the coupling between paraxial Bessel beams and trapped atoms. Based on this approach, we have derived simple analytical expressions for the excitation rates when considering linear, radial, and azimuthal polarization of incident light. While the theory is general and can be applied to any atom or ion, we have investigated the particular case of $^{2}S_{1/2} (F=0) \rightarrow\; ^{2}F_{7/2} (F=3, M=0)$ electric octupole transition in $^{171}$Yb$^{+}$. Our calculations have demonstrated that a proper choice of light polarization as well as of the mutual orientation of the beam direction and atomic quantization axis can strongly enhance this E3 transition. For a more accurate description of the excitation process we also took into account the effect of the atomic spatial distribution in a trap. It was found that the uncertainty in determination of atomic position significantly affects both the absolute value and the geometrical dependence of the transition rates. Thus, our study paves the way for understanding the optimal conditions for future experiments with trapped atoms and twisted photons.


%
%
%
%
\section*{\label{sec:acknowledgments}Acknowledgments}
We gratefully acknowledge the support of the Braunschweig International Graduate School of Metrology B-IGSM and the DFG Research Training Group 1952 Metrology for Complex Nanosystems. This research was also funded by the Deutsche Forschungsgemeinschaft (DFG, German Research Foundation) under Germany's Excellence Strategy - EXC-2123 QuantumFrontiers - 390837967 and within CRC 1227, Project S01.

\end{document}